\documentstyle[12pt]{article}
\renewcommand
\baselinestretch{1.4}

\def\beq{\begin{equation}}
\def\brr{\begin{array}}
\def\err{\end{array}}
\def\eeq{\end{equation}}
\def\bea{\begin{eqnarray}}
\def\eea{\end{eqnarray}}
\def\bs{\bigskip}

\def\ni{\noindent}
\def\wt{\widetilde}

\def\nn{\nonumber}
\def\ms{\medskip}

\begin{document}

\hfill HUPD-93-

\hfill UB-ECM-PF 93/

\hfill April 1993

\vspace*{3mm}

\begin{center}

{\LARGE \bf
The renormalization group and the effective potential in a curved
spacetime with torsion}

\vspace{4mm}

\renewcommand
\baselinestretch{0.8}
{\sc A.A. Bytsenko} \\ {\it Department of Theoretical Physics,
State
Technical
University, \\ St Petersburg 195251, Russia} \\
{\sc E. Elizalde}\footnote{E-mail address: eli @ ebubecm1.bitnet}
\\
{\it Department E.C.M., Faculty of Physics, University of
Barcelona, \\
Diagonal 647, 08028 Barcelona, Spain} \\
{\sc S.D. Odintsov}\footnote{On sabbatical leave from
Tomsk Pedagogical Institute, 634041 Tomsk, Russia.
E-mail address: odintsov @ theo.phys.sci.hiroshima-u.ac.jp} \\
{\it Department of Physics, Faculty of Science, Hiroshima
University, \\
Higashi-Hiroshima 724, Japan}
\ms

\renewcommand
\baselinestretch{1.4}

\vspace{5mm}

{\bf Abstract}

\end{center}

The renormalization group method is employed to study the
effective potential in curved spacetime with torsion. The
renormalization-group improved effective potential corresponding to
a massless gauge theory in such a spacetime is found and in this
way a generalization of Coleman-Weinberg's approach corresponding
to flat space is obtained. A method which works with the
renormalization group equation for two-loop effective potential
calculations in torsionful spacetime is developed. The effective
potential for the conformal factor in the conformal dynamics of
quantum
gravity with torsion is thereby calculated explicitly. Finally,
torsion-induced phase transitions are discussed.

\vspace{8mm}

\noindent PACS: 4.50, 3.70, 11.17.

\newpage

\section{Introduction}

It is a well-known fact that string theory contains an
antisymmetric two-form whose only dynamical content is that of an
axion [1]. The three-form gauge field strength corresponding to
this antisymmetric two-form may be interpreted as the torsion (for
a review of gravity with torsion see [2,3,15]). The ${\cal O}
(\alpha')$ string effective action [4] is equivalent to the usual
Einstein-Cartan theory, and to first order in $\alpha'$ it is
equivalent to higher-derivative gravity with torsion [5]. The
equivalence between an axion in string theory and the presence of
torsion has been discussed further in ref. [6], where it has been
shown also that black holes might have axion hair [6,7].

On the other hand, the current interest in gravity with torsion
[2,8] stems also from the search of the so-called fifth force
(for a review, see [4]). Moreover, to be noticed is also the fact
that cosmic strings can be naturally generated by torsion [9].

The present paper is devoted to the study of interacting quantum
field theory in curved spacetime with torsion. More specifically,
we apply the renormalization group approach to the calculation and
analysis of the effective potential corresponding to several
different theories.

The organization of the paper is as follows.
In the next section we develop the procedure which yields a
renormalization-group improvement of the effective potential
corresponding to an arbitrary massless gauge theory in a curved
spacetime with torsion. Some explicit examples are given. The
phase structure of the RG improved efective potential for the
$\lambda \varphi^4$ theory is discussed in detail. Sect. 3 is
devoted to the calculation of the two-loop effective potential in
a torsionful spacetime, in a situation in which all the
$\beta$-functions are known. In sect. 4 the same procedure of sect.
3  is applied in order to find the effective potential
corresponding to the
effective theory of the conformal factor in quantum gravity with
torsion. The phase transitions induced by torsion are discussed.
Such phase transitions can actually remove the original
singularity, a fact that might have very interesting consequences.
The conclusions of the paper are to be found in sect. 5.

\bs

\section{Renormalization-group improved effective potential in
curved spacetime with torsion}

In this section we  discuss the renormalization-group improved
effective potential for a theory in curved spacetime with torsion.
This will be done by generalizing Coleman-Weinberg's approach [10]
(see also [12]).
Let us consider a renormalizable, massless gauge theory which
includes scalars $\varphi$, spinors $\psi$ and gauge fields $A_\mu$
in a curved spacetime with non-zero torsion. We shall denote by
$\wt{g} \equiv (g,\lambda, h)$ the set of all coupling constants of
the theory ($g$ is the Yang-Mills, $\lambda$  the scalar and $h$
the Yukawa coupling), and $\wt{\xi}=( \xi, \zeta, \eta)$ are the
scalar-gravitational couplings. The tree level potential has the
following form
\beq
V^{(0)} = a\lambda \varphi^4-b\xi R \varphi^2-d\, \zeta S_\mu S^\mu
\varphi^2,
\label{2.1}
\eeq
where $a$, $b$ and $d$ are some positive constants.

The renormalization group equation for the effective potential has
the following form [11-15]
\beq
\left( \mu \frac{\partial}{\partial \mu} +\beta_{\wt{g}}
\frac{\partial}{\partial \wt{g}} +\delta \frac{\partial}{\partial
\alpha} +\beta_{\wt{\xi}} \frac{\partial}{\partial \wt{\xi}} -
\gamma \varphi \frac{\partial}{\partial \varphi} \right) V=0.
\label{2.2}
\eeq
We work in the Landau gauge ($\alpha =0$), in which $\delta =0$ to
one-loop order.

We adopt the approximation (linear on invariants of the
gravitational field) in which $\varphi^2 >> |R|$ and  $\varphi^2 >>
|S_\mu S^\mu|$. In this approximation we split $V$ in the following
way
\beq
V\equiv V_1+V_2+V_3 \equiv af_1(p,\varphi, \mu) \varphi^4- bf_2
(p,\varphi, \mu) R\varphi^2 - d\, f_3 (p,\varphi, \mu) S_\mu S^\mu
\varphi^2,
\eeq
where $p\equiv \{ \wt{g}, \wt{\xi}, \alpha \}$ and $f_1, f_2$ and
$f_3$ are some unknown functions. We also assume that each of the
three  $V_1, V_2$ and $V_3$ satisfy the renormalization group
equation (\ref{2.2}) (in this case, of course, $V$ satisfies
it too).

With all these considerations in mind, we can solve the
renormalization group equation (\ref{2.2}) as follows:
\beq
V= a \lambda (t) f^4(t) \varphi^4 -  b \xi (t) f^2(t) R\varphi^2
-  d\, \zeta (t) f^2(t) S_\mu S^\mu \varphi^2,
\label{2.4}
\eeq
where
\[
f(t)=\exp \left[ -\int_0^t dt'\, \bar{\gamma} \left( \wt{g} (t'),
\wt{\xi} (t'), \alpha (t') \right) \right], \ \ \ t=\frac{1}{2} \ln
\frac{\varphi^2}{\mu^2}, \ \ \ \dot{\wt{g}} (t) =
\bar{\beta}_{\wt{g}} (t), \]
\[ \dot{\wt{\xi}} (t) = \bar{\beta}_{\wt{\xi}} (t), \ \ \
\dot{\alpha} (t) = \bar{\delta} (t), \ \ \ \wt{g} (0) =  \wt{g}, \
\ \
\wt{\xi} (0) =  \wt{\xi}, \ \ \ \alpha (0) =\alpha \]
and
\[
\left( \bar{\beta}_{\wt{g}},   \bar{\beta}_{\wt{\xi}},
\bar{\gamma}, \bar{\delta} \right) = \frac{1}{1+ \gamma} \left(
\beta_{\wt{g}}, \beta_{\wt{\xi}}, \gamma, \delta \right). \]
The solution (\ref{2.4}) has been obtained using the following
initial conditions
\beq
V_1(t=0)= a\lambda \varphi^4, \ \ \ V_2(t=0)= -b\xi R \varphi^2, \
\ \   V_3(t=0)= -d\, \zeta S_\mu S^\mu \varphi^2.
 \eeq
The initial conditions for $V_1$ are slightly different from
Coleman-Weinberg's [10], what will lead to some differences in the
non-logarithmic terms; they are exactly the same as in ref. [13].
The renormalization-group improved effective potential
(\ref{2.4}) in the absence of torsion (i.e., $S_\mu =0$) has been
obtained in ref. [11].

In the one-loop approximation ---in which we shall actually work
throughout this paper--- the effective potential (\ref{2.4}) is
formally the same, but now with
\beq
f(t)=\exp \left[ -\int_0^t dt'\, \gamma \left( \wt{g} (t'),
\wt{\xi} (t') \right) \right], \ \ \dot{\wt{g}} (t) =
\beta_{\wt{g}} (t), \ \ \dot{\wt{\xi}} (t) = \beta_{\wt{\xi}} (t),
\ \  \wt{g} (0) =  \wt{g}, \ \ \  \wt{\xi} (0) =  \wt{\xi}.
\label{2.6}
\eeq
Expression (\ref{2.4}) with the functions of $t$ being given by
(\ref{2.6}) can be applied to a variety of gauge theories. We shall
now present a few examples.
\ms

\ni {\bf (i) $\lambda \varphi^4$-theory}. This is a quite simple
example, which may however be interesting enough from a pedagogical
point of view.
Here it is not necessary to have $\zeta \neq 0$ in order to obtain
multiplicative renormalizability. One can always put $\zeta =0$ and
then the theory does not interact with the torsion at all. Keeping,
though,  $\zeta \neq 0$, and using the well-known values for the
effective coupling constants, we get
\beq
V= \frac{\lambda \varphi^4}{4!\, \left( 1- \frac{3\lambda t}{(4\pi
)^2}\right) } - \frac{1}{2} R\varphi^2 \left[ \frac{1}{6} + \left(
\xi - \frac{1}{6} \right) \left( 1- \frac{3\lambda t}{(4\pi)^2}
\right)^{-1/3} \right] - \frac{1}{2} S_\mu S^\mu \varphi^2 \left[
\zeta  \left( 1- \frac{3\lambda t}{(4\pi)^2} \right)^{-1/3}
\right],
\label{2.7}
\eeq
where $t=\frac{1}{2} \ln (\varphi^2 /\mu^2)$. The potential
(\ref{2.7}) is valid for the range of values of $t$ which make it
not to diverge (in particular, for any negative value of $t$). In
flat space ($R=S_\mu =0$) the effective potential (\ref{2.7}) has
been obtained in [10], and in curved space without torsion ($R\neq
0$, $S_\mu=0$), in [11].
\ms

\ni {\bf (ii) Asymptotically free SU(2) theory}. This is an example
in which torsion appears in a much more natural way than in the
previous case. The Lagrangian is given by (see [14])
\bea
L &=& -\frac{1}{4}  G_{\mu\nu}^a G^{a\mu\nu} + \frac{1}{2}
g^{\mu\nu} (D_\mu \varphi )^a  (D_\nu \varphi )^a +  \frac{1}{2}
\xi R \varphi^a \varphi^a  +  \frac{1}{2} \zeta S_\mu S^\mu
\varphi^a \varphi^a \nn \\
&-& \frac{1}{4!}\lambda (\varphi^a \varphi^a)^2 + i \bar{\psi}^a
\left[
\gamma^\mu D_\mu^{ab} + \eta \gamma_5 \gamma^\mu S_\mu \delta^{ab}
\right] \psi^b - i h \epsilon^{acb}  \bar{\psi}^a \varphi^c \psi^b.
\eea
Here $ (D_\mu \varphi )^a = \partial_\mu \varphi^a + g
\epsilon^{abc}  A_\mu^b \varphi^c$ and $ D_\mu^{ab} \psi^b =
\nabla_\mu \psi^a  + g  \epsilon^{abc}  A_\mu^b \varphi^c$, the
gauge group is  SU(2), and $\varphi^a$ and $\psi^a$ belong to the
adjoint representation of the gauge group. It is known that the
theory under discussion is asymptotically free for special
solutions of the reormalization group [14].

It is not difficult to show that the theory minimally coupled to
torsion and metric ($\eta =-1/8$, $\xi = \zeta =0$) is {\it not}
multiplicatively renormalizable [14] (for a general discussion see
[15]). In order to get multiplicative renormalizability, we must
introduce the coupling parameters $\xi$, $\zeta$ and $\eta$ [15].

The renormalization group equations for the effective couplings
have been obtained in refs. [14]:
\bea
&& g^2(t)= \frac{g^2}{1+l^2t}, \ \ l^2= \frac{b^2g^2}{(4\pi )^2},
\ \ b^2= \frac{26}{3}, \ \ \lambda (t) = \kappa_1 g^2(t), \ \
\kappa_1 = \sqrt{\frac{97}{22}}, \ \ h^2 (t) = \kappa_2 g^2(t), \nn
\\
&& \kappa_2 = \frac{23}{24}, \ \ \xi (t) = \frac{1}{6} + \left( \xi
- \frac{1}{6} \right) (1+l^2 t )^{-a^2/b^2}, \ \ a^2= 12- \frac{5
}{3} \kappa_1 - 8 \kappa_2 >0, \nn  \\ && \eta (t) = \eta  (1+l^2
t )^{4\kappa_1^2/b^2}, \ \ \
 \zeta (t) = \left( \zeta + \frac{32\eta^2}{8\kappa_2+b^2}
\right) (1+l^2 t )^{-a^2/b^2} \nn \\ &&  -
\frac{32\eta^2}{8\kappa_2+b^2}  (1+l^2 t )^{8\kappa_1/b^2},
\ \ f(t)= (1+l^2 t )^{(6-4\kappa_2)/a^2}.
\label{2.9}
\eea
Using the effective coupling constants (\ref{2.9}), we obtain the
renormalization-group improved effective potential
\beq
V=\frac{1}{4!} \lambda (t) f^4(t) \varphi^4 - \frac{1}{2} \xi (t)
R
f^2(t) \varphi^2 - \frac{1}{2} \zeta (t) S_\mu S^\mu  f^2(t)
\varphi^2,
\label{2.10}
\eeq
where $\varphi^2 = \varphi^a \varphi^a$ and $t=\frac{1}{2} \ln
(\varphi^2 /\mu^2)$. In the same way, one can easily get the
effective potential for a variety of models (for example, GUTs, see
ref. [15]), where the effective coupling constants are known.
Let us now investigate the possiblity of a first order phase
transition induced by the external gravitational field (the
possibility of a phase transition induced by torsion has been
already pointed out in ref. [16]). A detailed analysis of
curvature-induced phase transitions has been done in our previous
work, ref. [11].

With torsion the situation changes as follows. For simplicity of
the discussion, let us first restrict ourselves to the case of the
$\lambda \varphi^4$ theory, although our considerations also apply
to the SU(2) case and are indeed very general. As is manifest from
the specific form of the potential (\ref{2.7}) (and also from that
of the potential (\ref{2.10})), the inclusion of torsion turns out,
in the end, in the
appearance of an additional term which has exactly the same
$t$-dependence as the main term corresponding to non-zero
curvature. This is to say, in the presence of both torsion and
curvature, the analysis of phase transitions is basically the same
as the analysis corresponding to curvature alone: no specifically
new situation is created by the addition of torsion.

However, the remarkable thing is the fact that also in the absence
of curvature, the presence of torsion still reproduces some of the
behaviors typically induced by curvature. This is true both for
the  $\lambda \varphi^4$ theory and also for the SU(2) one.
The potential (\ref{2.7}) can be written in that case as
\beq
y= \frac{1}{2u(x)} \left[ \frac{\lambda}{12} \, x^2 -
\frac{\zeta}{\mu^2} u(x)^{2/3} S^2 \, x \right],
\label{y11}
\eeq
where
\beq
x\equiv \frac{\varphi^2}{\mu^2}, \ \ \ \ y\equiv \frac{V}{\mu^4},
\ \ \ \ u(x) \equiv 1- \frac{3\lambda \ln x}{2(4\pi)^2}.
\eeq
A direct analysis of the parenthesis in (\ref{y11}) shows that it
just has two different possible behaviors, as a function of $x$.
Namely, for
positive $\zeta$, it is a function that starts at the origin, goes
down and reaches a minimum value at some $x_m\neq 0$, and ends as
an increasing parabolic-like curve (Fig. 1a); on the other hand,
for negative $\zeta$
the minimum is obtained at the origin and the curve is
monotonically increasing all the time (Fig. 1b).

\bs
\section{Two-loop effective potential in curved spacetime with
torsion}
We shall here develop a method for the calculation of the massless
effective potential at {\it any} loop order. The method has its
roots in the
direct solution of the RG equations for a massless gauge theory in
curved spacetime with torsion. (Again, we are going to work in the
approximation where the invariants of the gravitational field
---$R$ and $S_\mu S^\mu$--- appear linearly). This method has been
developed already for flat space [17] and generalized later to
curved
spacetime [18]. Here we will give the closed expressions
corresponding to the two-loop effective potential. As an example of
two-loop effective potential calculation, that for the $\lambda
\varphi^4$-theory in curved spacetime with torsion will be carried
out explicitly.

Starting once more with the effective potential (\ref{2.1}), we
consider again the RG equation (\ref{2.2}) in the Landau gauge.
Working in the same approximation as in sect. 2, we shall  also
assume that each of the $V_1$, $V_2$ and $V_3$ satisfy eq.
(\ref{2.2}) independently. (Then, of course, $V$ will also satisfy
it). Using (\ref{2.1}), we can write the following $n$-loop order
RG
equation for the effective potential
\beq
 \mu \frac{\partial}{\partial \mu} V^{(n)} + D_n V^{(0)}+  D_{n-1}
V^{(1)}+ \cdots + D_1 V^{(n-1)}=0,
\label{3.1}
\eeq
where $V^{(n)}$ is the $n$-loop correction to the effective
potential,
\[
D_n = \beta_{\wt{g}}^{(n)} \frac{\partial}{\partial \wt{g}}
 +\beta_{\wt{\xi}}^{(n)} \frac{\partial}{\partial \wt{\xi}} -
\gamma^{(n)} \varphi \frac{\partial}{\partial \varphi},
\]
and $\beta^{(n)}$ is the $n$-loop correction to the corresponding
$\beta$-function. In accordance with our proposal, we have three
equations (\ref{3.1}) ---for  $V_1$, $V_2$ and $V_3$--- and we can
find the effective potential by using the recursion formula
(\ref{3.1}). For $V_1$ and $V_2$ this has been done already in
refs. [17,18], respectively, where a closed expression up to
two-loop order has been obtained. $V_3$ can be found in a similar
way. Using the following renormalization conditions
\beq
\left. V_i^{(j)} \right|_{\mu = \varphi} =0, \ \ \ \ i=1,2,3, \ \
j=1,2,
\eeq
we get
\bea
V&=& V^{(0)} + V^{(1)} + V^{(2)} = a\lambda \varphi^4 + A^{(1)}
\varphi^4 \ln \frac{\varphi^2}{\mu^2} + \frac{1}{2} \left[ \left(
\beta_\lambda^{(2)} -4\lambda \gamma^{(2)} \right) a- 2
\gamma^{(1)} A^{(1)} \right] \varphi^4 \ln \frac{\varphi^2}{\mu^2}
\nn \\
&+& \frac{1}{4} \left[ \beta_{\wt{g}}^{(1)} \, \frac{\partial
A^{(1)}}{ \partial \wt{g}} -4 \gamma^{(1)} A^{(1)} \right]
\varphi^4 \left( \ln \frac{\varphi^2}{\mu^2} \right)^2 -b\xi R
\varphi^2 - B^{(1)} R \varphi^2 \ln \frac{\varphi^2}{\mu^2} \nn \\
&-&  \frac{1}{2} \left[  \beta_\xi^{(2)} -2\xi \gamma^{(2)} -
 2 \gamma^{(1)} \frac{B^{(1)}}{b} \right] bR \varphi^2 \ln
\frac{\varphi^2}{\mu^2} - \frac{1}{4} \left[ \beta_{\wt{g}}^{(1)}
\, \frac{\partial B^{(1)}}{ \partial \wt{g}}+
\beta_{\wt{\xi}}^{(1)} \, \frac{\partial B^{(1)}}{ \partial
\wt{\xi}} -2 \gamma^{(1)} B^{(1)} \right] \nn \\ &\times & R
\varphi^2
\left( \ln \frac{\varphi^2}{\mu^2} \right)^2
- d \, \zeta S^2 \varphi^2- D^{(1)} S^2 \varphi^2 \ln
\frac{\varphi^2}{\mu^2} - \frac{1}{2} \left[  \beta_\zeta^{(2)} -
2\zeta \gamma^{(2)}       - 2 \gamma^{(1)} \frac{D^{(1)}}{d}
\right] d\, S^2 \varphi^2 \ln \frac{\varphi^2}{\mu^2} \nn \\
&-&  \frac{1}{4} \left[ \beta_{\wt{g}}^{(1)} \, \frac{\partial
D^{(1)}}{ \partial \wt{g}}+  \beta_{\wt{\xi}}^{(1)} \,
\frac{\partial D^{(1)}}{ \partial \wt{\xi}} -2 \gamma^{(1)} D^{(1)}
\right] R \varphi^2 \left( \ln \frac{\varphi^2}{\mu^2} \right)^2,
\label{3.3}
\eea
where
\[ A^{(1)} =\frac{a}{2} \left( \beta_\lambda^{(1)} -4\lambda
\gamma^{(1)} \right), \ \  B^{(1)} =\frac{b}{2} \left(
\beta_\xi^{(1)} -2\xi \gamma^{(1)} \right), \ \   D^{(1)}
=\frac{d}{2} \left( \beta_\zeta^{(1)} -2\zeta \gamma^{(1)} \right).
\]
Using (\ref{3.3}) one can immediately obtain the two-loop effective
potential for any gauge theory in a spacetime with torsion.

As an example, we give now the result corresponding to the $\lambda
\varphi^4$-theory. The two-loop beta functions for flat spacetime
have been obtained in ref. [19] already, while the two-loop
$\beta_\xi$ has been calculated in ref. [20]; moreover, up to two
loops, we have $\beta_\zeta =\zeta \gamma_{m^2}$ [15], and the
two-loop $\gamma$-function for the scalar field mass is given, for
instance, in [19]. Using these $\beta$-functions, we get
\bea
V &=& \frac{\lambda}{24} \varphi^4 - \frac{\xi}{2} R \varphi^2 -
\frac{\zeta}{2} S^2 \varphi^2 + \frac{\lambda^2}{(16\pi)^2}
\varphi^4  \ln \frac{\varphi^2}{\mu^2}- \frac{\lambda}{(8\pi)^2}
\left( \xi - \frac{1}{6} \right) R \varphi^2 \ln
\frac{\varphi^2}{\mu^2} -
 \frac{\lambda \zeta S^2}{(8\pi)^2}  \varphi^2 \ln
\frac{\varphi^2}{\mu^2} \nn \\  &-&  \frac{\lambda^3}{8(4\pi)^4}
\varphi^4
\ln \frac{\varphi^2}{\mu^2} +  \frac{3\lambda^3}{32(4\pi)^4}
\varphi^4 \left( \ln \frac{\varphi^2}{\mu^2} \right)^2 -
\frac{\lambda^2}{4(4\pi)^4} \left[ \left( \xi - \frac{1}{6}
\right)+ \frac{1}{36} \right] R \varphi^2 \ln
\frac{\varphi^2}{\mu^2} \nn \\
&-& \frac{\lambda^2}{4(4\pi)^4} \left( \xi - \frac{1}{6} \right) R
\varphi^2 \left( \ln \frac{\varphi^2}{\mu^2} \right)^2 -
\frac{\lambda^2}{4(4\pi)^4} \zeta S^2 \varphi^2 \ln
\frac{\varphi^2}{\mu^2}  - \frac{\lambda^2}{4(4\pi)^4} \zeta
S^2 \varphi^2 \left( \ln \frac{\varphi^2}{\mu^2} \right)^2.
\label{3.4}
\eea
In more realistic theories, as the SU(2) model, the analog of eq.
(\ref{3.4}) is actually more interesting because the contribution
of the fermion coupling $\eta$ will appear in terms connected with
the torsion (even in the case of minimal coupling). The final
expressions are indeed very complicated and, moreover,
$\beta_\zeta$ to two loops is not known in such theories. Notice
also that, as it follows from (\ref{3.4}), for zero curvature but
non-zero torsion and $\zeta \neq 0$, we obtain spontaneous symmetry
breaking induced by torsion. At tree level and for $\zeta >0$, we
get
\beq
\varphi^2 =\frac{6\zeta S^2}{\lambda}.
\label{3.5}
\eeq
Using (\ref{3.4}) we can then obtain loop corrections to the
minimum (\ref{3.5}).

To summarize, we have proven in this section that the method
developed allows us, in fact, to calculate the multiloop effective
potential for a field theory in any spacetime with torsion.
\bs

\section{Phase transitions induced by torsion in infrared quantum
gravity}

In this section we will study the effective potential which arises
in the conformal dynamics of quantum gravity with torsion [22] (for
a discussion of conformal dynamics of quantum gravity, see [21]).

Let us first briefly recall the construction of the trace anomaly
induced dynamics of the conformal factor [22]. We start from the
free, conformally invariant theory corresponding to $N_0$ scalars,
$N_{1/2}$ spinors and $N_1$ vectors on a spacetime with torsion
(see [15]). It is given by the action
\beq
S=S_0+ S_{1/2} + S_1,
\label{4.1}
\eeq
where
\bea
S_0 &=& \frac{1}{2} \int d^4x \, \sqrt{-g} \left[ g^{\alpha\beta
}\partial_\alpha \varphi \partial_\beta \varphi +\frac{1}{6}
R\varphi^2 + \zeta S_\mu S^\mu \varphi^2 \right], \nn \\
S_{1/2} &=& i \int d^4x \, \sqrt{-g} \left[ \bar{\psi} \left(
\gamma^\mu \nabla_\mu - \eta \gamma_5 \gamma^\mu S_\mu \right) \psi
\right], \nn \\
S_1&=& -\frac{1}{4} \int d^4x \, \sqrt{-g} \, G^2_{\mu\nu},
\eea
being $\zeta$ and $\eta$ arbitrary coupling constants. The
theory with the action (\ref{4.1}) is conformally invariant for any
value of  $\zeta$ and $\eta$. The minimal coupling corresponds to
 $\zeta =0$ and $\eta =1/8$.

The trace anomaly for the theory (\ref{4.1}) in curved spacetime
with torsion is given by the following expression (see [15] for
details and references):
\bea
T_\mu^\mu &=& b C^2_{\mu\nu\alpha\beta} + b' \left( G- \frac{2}{3}
\Box R \right) + \left[ b'' + \frac{2}{3} (b+b'')  \right] \Box R
+ a_1F^2_{\mu\nu} \nn \\
&+& a_2 (S_\mu S^\mu)^2 + a_3 \Box (S_\mu S^\mu) + a_4 \nabla_\mu
\left(S_\nu \nabla^\nu S^\mu- S^\mu  \nabla_\nu S^\nu \right),
\eea
where $G$ is the Gauss-Bonnet invariant, $F_{\mu\nu} = \nabla_\mu
S_\nu-  \nabla_\nu S_\mu$, and where the coefficients $b,b',
\ldots, a_4$
are well known (see, for instance, [21,22]). In particular, the
coefficients $a_1, \ldots, a_4$ relevant for non-zero torsion are
\bea
&& a_1 =- \frac{2}{3(4\pi)^2} \Sigma \eta^2, \ \ \ \ a_2=
\frac{1}{2(4\pi)^2} \Sigma \zeta^2, \nn \\
&& a_3 = \frac{1}{3(4\pi)^2} \Sigma \left( 2\eta^2- \frac{1}{2}
\zeta^2 \right), \ \ \ \ a_4= - \frac{2}{3(4\pi)^2} \Sigma \eta^2.
\eea

Choosing the conformal parametrization
\beq
g_{\mu\nu} = e^{2\sigma (x)} \eta_{\mu\nu}, \ \ \ \ S_\mu =
\bar{S}_\mu,
\label{4.5}
\eeq
where $\sigma$ is the conformal factor, $\eta_{\mu\nu}$ the
Minkowski metric and $\bar{S}_\mu$ an arbitrary constant torsion
background, one can integrate over the trace anomaly in order to
get the trace-anomaly-induced effective action, $S_{anom}$ (see
[15] for details). Adding the classical gravity action
\beq
S_{cl}= \frac{1}{2\kappa} \int d^4x \, \sqrt{-g} \, \left( R+ h
S_\mu S^\mu -2\Lambda \right),
\eeq
with the parametrization (\ref{4.5}) to $S_{anom}$, we get the
total effective action which describes the conformal factor
dynamics [22]
\bea
S_{eff} &=& S_{anom}+ S_{cl} = -\frac{Q^2}{(4\pi)^2} \int d^4x\,
(\Box \sigma )^2- \zeta \int d^4x \, \left[ 2\alpha (\partial_\mu
\sigma)^2 \Box \sigma + \alpha^2   (\partial_\mu \sigma)^4 \right]
\nn \\
&+& \gamma  \int d^4x \, e^{ 2\alpha \sigma} (\partial_\mu
\sigma)^2 - \frac{\lambda}{\alpha^2}  \int d^4x \, e^{ 4\alpha
\sigma}+   \int d^4x \, \left[ \left( a_3 + \frac{a_4}{2} \right)
\bar{S}^2  (\partial_\mu \sigma)^2 \right. \nn \\
&+& \left. a_4 \bar{S}^\mu \bar{S}^\nu \partial_\mu \sigma
\partial_\nu \sigma + \frac{h}{2\kappa \alpha^2}  e^{ 2\alpha
\sigma} \bar{S}^2 \right],
\eea
where the transformations $\sigma \rightarrow \sigma \alpha$ and
$S_{eff} \rightarrow \alpha^{-2} S_{eff}$ have been performed, and
\beq
\frac{Q^2}{(4\pi)^2} = 2b+3b', \ \ \ \zeta = 2b+2b'+3b'', \ \ \
\gamma = \frac{3}{\kappa}, \ \ \ \lambda = \frac{\Lambda}{\kappa}.
\label{4.7}
\eeq
In what follows we are going to work about the infrared stable
fixed point $\zeta =0$ (see [21,22]). Let us denote
$e^{\alpha\sigma} \equiv \Phi$. (Notice that $\Phi$ is always
positive).

The tree level effective potential is given by
\beq
V(\Phi) = \frac{\lambda}{\alpha^2} \Phi^4 -   \frac{h}{2
\kappa\alpha^2} \Phi^2 S^2.
\eeq
The one-loop correction to this potential can be easily found using
the general expression (\ref{3.3}) (the one-loop beta functions are
known from [21,22]). The result is (we use Coleman-Weinberg's
normalization conditions)
\bea
V^{(1)} (\Phi )&=&  \frac{\lambda}{\alpha^2} \Phi^4 - \left[
\frac{\gamma^2 (4\pi)^2}{4Q^4}-\frac{4\lambda}{Q^2} \right] \Phi^4
\left( \ln \frac{\Phi^2}{\mu^2} - \frac{25}{6} \right)- \frac{h}{2
\kappa\alpha^2} \Phi^2 S^2 \nn \\
&+&  \left[ \frac{\gamma (4\pi)^2}{2Q^4} \left(a_3+ \frac{3}{4}
a_4\right) -\frac{h}{2\kappa Q^2} \right] \Phi^2 S^2 \left( \ln
\frac{\Phi^2}{\mu^2} - 3 \right).
\label{4.9}
\eea
As we can see, in the absence of torsion the minimum corresponding
to the tree effective potential is obtained for
\beq
\Phi =0, \ \ \ \ \ \ \sigma \longrightarrow -\infty.
\eeq
In terms of the original metric, this corresponds to the
singularity. However, the remarkable point is the fact that
symmetry breaking appears in two different ways: already at tree
level as a result of adding torsion, or else at one-loop because of
quantum corrections (see also [23]). Both lead to a non-zero
vacuum, which in the second case is (for simplicity, we take
$S^2=0$)
\beq
\frac{\sigma}{\sigma_0} =- \frac{\lambda}{\alpha^2}  \left[
\frac{\gamma^2
(4\pi)^2}{4Q^4}-\frac{4\lambda}{Q^2} \right]^{-1}
+\frac{25}{6}-\frac{11}{4\alpha}
\eeq
(the parameters should be chosen to have on the right-hand side the
positive ones).
Hence, the singular vacuum becomes the nonsingular one as a result
of the Coleman-Weinberg symmetry breaking (the singularity is
avoided). In principle one can construct the renormalization group
improved effective potential as in sect. 2. Nevertheless, in the
present case we have many coupling constants and the corresponding
effective coupling constants do not show such a simple behavior as
before (like asymptotic freedom, for example). For this reason we
shall not discuss the RG improved effective potential here.

Let us now investigate the possibility of a phase transition
induced
by torsion in the effective conformal factor theory (\ref{4.7}).
It turns out that an exact analytical study can again be carried
out. It
leads to the following results. To start with, eq. (\ref{4.9}) can
be written in the shortened form
\beq
y=F(x)= x^2+ax^2 \left( \ln x -\frac{25}{6} \right) -bx+ c x (\ln
x -3), \ \ \ \  x>0,
\eeq
where the $x$, $y$  and the constants $a$, $b$ and $c$ are
immediately identified by simple inspection
\bea
&& x \equiv \frac{\Phi^2}{\mu^2}, \ \ \ \ y\equiv
\frac{\alpha^2}{\lambda\mu^4} \, V(\Phi ), \ \ \ \ a\equiv
\frac{\alpha^2}{Q^2} \left[ \frac{\gamma^2 (4\pi)^2}{4Q^2}-4\lambda
\right], \nn \\
&& b\equiv \frac{h}{2\lambda \kappa \mu^2} \, S^2, \ \ \ \ c \equiv
\frac{\alpha^2}{2\lambda \mu^2 Q^2} \left[ \frac{\gamma
(4\pi)^2}{Q^2} \left(a_3+ \frac{3}{4}
a_4\right) -\frac{h}{\kappa } \right]  S^2.
\eea
Notice again that $x>0$ and that $b\geq 0$. Proceeding with the
calculation of extrema, the first derivative yields
\beq
y'= (2ax+c) \left[ \ln x + \left( \frac{1}{a}- \frac{11}{3} \right)
+ \frac{ \frac{5}{3}c-\frac{c}{a}-b}{2ax+c} \right].
\eeq
 Thus, the
extrema are obtained as the crossing points of the two functions
\beq
\ln x = G(x),   \ \ \ \ G(x)\equiv  \left[  \left( \frac{11}{3}-
\frac{1}{a} \right) + \frac{ -\frac{5}{3}c+\frac{c}{a}+b}{2ax+c}
\right].
\label{root2}
\eeq
It is immediate that the function $G(x)$ is either monotonically
increasing in the whole range $x>0$ or else monotonically
decreasing in the whole range (its first derivative has a constant
sign). For $x \rightarrow \infty$ it goes
asymptotically to the constant value $G(\infty )= \frac{11}{3}-
\frac{1}{a}$, unless
$a\equiv 0$, in which case it is constantly equal to $G(x) \equiv
2+\frac{b}{c}$, and a single extremum is obtained. This last
value is, in general, the one reached at the origin $G(0)$.
The convexity of $G(x)$ has also a uniform sign in the whole range
or can, at most, change sign once. All this leads to the conclusion
that eq.
(\ref{root2}) has at most two solutions. The  two extrema
can be obtained either both of them at the same side of the
discontinuity of $G(x)$ (i.e., $x_0=-c/2a$) or each at one side of
it. We shall be specially interested in the case $a>0$ and $c<0$,
where the usual mexican-hat shape appears (Figs. 2a and 2b).
However, the
phase transition ---induced by torsion--- is obtained  for a very
wide range of values of the parameters. In fact, it already shows
up for very small values of $\alpha^2$ and of the torsion $S^2$.
Some results, with the typical form of the symmetry breaking
potential, corresponding to several different values of the
constants, are depicted in Fig. 2. The final stages (Figs. 2c and 2d)
look the same as in Fig. 1. Similar questions have been investigated
very recently for a different model in ref. [24].
\bs

\section{Conclusions}

We have developed in this paper a formalism which yields a
well-defined procedure to study the effective potential, and the
corresponding phase structure, of gauge theories in torsionful
spacetime. In particular, the
renormalization-group improved effective potential for any massless
gauge theory has been thus found. That is, we have been able to
extend the one-loop effective potential formalism to a spacetime
with torsion,
taking thereby into account all logarithmic corrections. There is
no problem, in principle, to extend our approach to massive
theories (except for some minor complications of technical nature).
This will be done elsewhere.

Moreover, a method, based on the RG, for the calculation of the
corresponding multiloop
effective potential in a torsionful spacetime, has been also
constructed,
thus generalizing again the corresponding method valid for flat
space.

Finally, the effective potential corresponding to the conformal
sector of quantum
gravity with torsion has been discussed. It has been shown that,
for different values of the parameters of the theory, a phase
transition induced by torsion may take place. In particular, this
phase transition might lead to a removal of the original
singularity. Such phenomenon could be very relevant to early
Universe considerations.

\vspace{5mm}

\ni{\large \bf Acknowledgments}

S.D.O. wishes to thank I. Antoniadis, F. Englert and T. Muta for
discussions
on related topics and the Particle Group at Hiroshima University
for kind hospitality.
S.D.O. has been supported by JSPS (Japan) and
E.E.  by DGICYT (Spain), research project
PB90-0022.
\bs



\newpage

\ni{ \large \bf Figure captions}
\bs \bs

\ni{\bf Figure 1.} The renormalization-group improved effective
potential ($y=V/\mu^4$) corresponding to the $\lambda \varphi^4$
theory
in a curved spacetime with torsion, eqs. (7) and (11), as a
function of
$x=\varphi^2/\mu^2$. In Fig. 1a, $\zeta$ is positive. In Fig. 1b,
$\zeta$ is negative.
\bs

\ni{\bf Figure 2.} The function $y=F(x)$, eqs. (30) and (27),
corresponding
to the effective potential for infrared quantum gravity with
torsion for
different values of the constants (31). All them have been taken
between 0  and 1, except for $c$ in Figs. 2a and 2b, which is
negative
and of order 10. When torsion varies, the typical evolution of the
symmetry breaking potential is obtained.

\end{document}